\documentclass[11pt]{article}
\usepackage[usenames]{color}
\begin{document}
\sloppy
\author{\textbf{Arbab I. Arbab}\footnote{On leave from Comboni College, P.O. Box 114, Khartoum,
Sudan, E-mail: arbab@ictp.trieste.it}\\ \\
Department of Physics, Teachers' College, \\
P.O. Box 4341, Riyadh 11491, Saudi Arabia }
\title{\bf Quantization of Gravitational System and its Cosmological Consequences}
\maketitle
\begin{abstract}
We have found that the hierarchial problems appearing in cosmology
is a manifestation of the quantum nature of the universe. The
universe is still described by the same formulae that once hold at
Planck's time. The universe is found to be governed by the Machian
equation, $G M=Rc^2$, where $M$ and $R$ are mass and radius of the
universe. A Planck's constant for different cosmic scales is
provided. The  status of the universe at different stages is shown
to be described in terms of the fundamental constants (eg.
\textcolor{blue}{$c, \hbar, G, \Lambda, H$}) only. The concept of
maximal (minimal) acceleration, power, temperature, etc., is
introduced and justified. The electromagnetic interactions are
shown to be active at a cosmic level. Their contribution would
exclude the inclusion of dark energy in cosmology.
\end{abstract}
\vspace{1cm} \textbf{Key Words}: {Cosmology: early universe,
cosmological parameters, quantum gravity} \vspace{.2in}
\baselineskip=19pt
\section{\textcolor{blue}{Introduction}}
Einstein  added the cosmological constant to his general
relativity equations in an attempt to get a gravitationally stable
universe. When he later knew that the universe is expanding
dropped his constant and regretted its addition. Cosmologists and
particle physics have found that this term is connected with a
constant vacuum (minimum) energy of the spate-time [1, 2, 3]. The
value of this minimum energy is estimated to be very large in the
early universe. However, at the present time the value of this
constant is found to be vanishingly small. This contradiction is
now coined in the term ``cosmological constat problem". To solve
this problem several models [1,2, 3, 4] are proposed. Some models
considered this constant to decay with time (or scale factor) [5,
6, 7]. These models are now known as vacuum decaying models.
Others relate this to the energy density of some scalar field
dominating the universe at early times having some properties that
gives the present observed value. Scientists have been  looking
for some symmetry to set it to zero. In some supersymmetry models
[4] it cancels but when the supersymmetry is broken to give the
present universe its value becomes non-zero.

In fact, the present unresolved problems in cosmology has to do
with the non-existence of a complete quantum gravity which blends
general relativity with quantum mechanics. There are however, some
hopes in the string theory to come up with the final remedy.
General relativity is successful in describing large scale
structures (stars, galaxies, universe), while quantum mechanics is
applicable to microscopic scale(atomic). However, when quantum
mechanics applied to macroscopic objects it doomed with failure.
It seems there is no intersection region in which both systems
converge. There is however some cases in which quantum mechanics
was successful when applicable to cosmic systems, e.g., black
holes. Thus it might be still possible to apply such a
method to other gravitational systems. \\
The point is that when people apply quantum mechanics to large
scale systems, they use the same unit of quantization,i.e., the
Planck's constant. In fact, Planck's constant is a measure of the
precision in which physical quantities in quantum mechanics can be
distinguished. A microscopic system has an angular momentum of the
order of Planck's constant. But one does not expect, for instance,
a Sun to have such a small value. Therefore, one does not
anticipate that this value to be so small for a macroscopic
system. Contrarily, one would expect a macroscopic system to have
a characteristics unit of quantization (a cosmic Planck's
constant) that is so huge. For instance, the spin angular momentum
of the Earth is $\sim 10^{34}\rm J s$.

We know that a quantum formula approximates to the classical
counterpart when  $\hbar\rightarrow 0$. However, we do see that in
classical world the unit of angular momentum is so big, so we
expect an analogous quantum formula  for macroscopic system to
hold when $\hbar\rightarrow\infty$ instead, while all other
quantities scaled up (mass, distance, etc). Moreover, we argue
that cosmic Planck's constant is not unique but assumes different
values from one system to another. With this idea in mind, we have
found that when appropriate arrangements are made, quantum
mechanical formulae of microscopic system apply equally well to
macroscopic system. One of the consequences of this conjuncture is
that evolution of black holes. It is remarkable that the
Schwarzschild radius of a black hole emerges naturally from this
conjecture of cosmic quantization; it corresponds to the Compton
(or de Broglie) wave length defined in terms of this cosmic
Planck's constant. As we know, when the dimension  of a physical
system becomes of the order of Compton (de Broglie) wavelength,
quantum effects become prominent. We thus see that quantum
mechanics can be applied to microscopic systems as well as
macroscopic ones, i.e., from a very small scale to a very large
scale. This formalism shares the concept of duality manifested by
string theories\footnote{\textcolor{blue}{See, J. Polchinski,
Rev. Mod. Phys. 68 (1996) 1245}}. We have found the cosmic
Planck's constant ($\hbar_c$) is scaled as $\hbar_c\propto
(\frac{1}{\hbar})^n$, for some $n$. Such a framework may be linked
with some kind of a scale invariance of the theory of quantum
gravity.

One is inclined to ask the question whether  the condition
$\hbar\rightarrow 0$ is the only way to go from quantum theory to
the classical theory? Had we adopted another formalism of quantum
mechanics, would it be possible for us to find a correspondence
law in which $\hbar\rightarrow \infty$ gets us back the classical
regime? Or would it be possible to have a classical formalism
which allows us to go from classical to quantum world by setting,
say Newton's constant ($G) \rightarrow 0$\ ? In this circumstance,
one will have a bi-pass joining  classical and quantum pictures of
the world. This will help us that whenever advances are made in
classical world a quantum mechanical analogue has to be assumed.
This may seem to be plausible owing to the similarity between
gravity and electromagnetism. There might be some symmetry which
can render this transformation successful. In such a case one
would say that a quantum system has more than one classical limit.
Or reversely, one can go from the classical system to a quantum
system by applying the inverse transformation? If such a duality
existed, the description of natural laws becomes more interesting.
One asks a question that, will it be possible to go from
gravitational system to electromagnetic system owing to the
similarity of both system? Could one use the relation $e^2=Gm^2$
?; and using the known formulae of the electromagnetic system and
apply it to the classical system. In this situation one finds that
the fine structure constant $\alpha_E\rightarrow
\alpha_G=\frac{Gm^2}{\hbar c}$, leading to $\alpha_E=\alpha_G$ (if
one uses $m=m_{\rm Planck}=1.85\times 10^{-6}\rm
g$)\footnote{\textcolor{blue}{We obtain this value from the
relation $e^2=Gm^2$ and substituting for $e$ its value in esu.}}.
Hence, gravity gets unified with electromagnetism at Planck's
time. However, according to our conjuncture of cosmic
quantization, one finds for the whole universe (today) that
$\alpha_G=1$. That means that gravity becomes a strong force at
large scale while the other forces should be compared with it.
This would imply that at the cosmic level, electromagnetism and
strong forces become weak in comparison with gravity. This is the
opposite situation for nuclear (or atomic) interactions. Thus, one
can transform gravitational system from being very weak to be very
strong. Such a similarity principle may hold between
electromagnetic and gravitational system. Thus one sees that these
points should be considered before a successful theory of quantum
gravity being endeavored. If this works well, we should make some
critical changes to our understanding of the physical laws
describing our world. One would hope that string theory can
provide such a mechanism (or a transformation) by finding a
formalism that realizes this idea. If this is realized, then one
would argue that the classical system is a real quantized system
if correctly read.

A proper unification of all interactions should include the
fundamental constants, representing the four basic interactions
($G, c, \hbar, \Lambda$), in its premises. An appropriate
combinations of these constants can provide a proper description
of the anticipated unified interaction. We have found that these
fundamental constants describe completely our universe, at all
stages.

Such a prescription may help elucidate the way to a fully quantize
gravity. Aided with this belief a lot of the cosmological problem
can be alleviated, as we will see here. In this work we have seen
that the present contribution of the vacuum energy in no longer a
puzzle. This is because this vacuum energy evolves from Planck's
(quantum) value to its present value as a result of the
quantization of the cosmological parameters, and that our vacuum
energy is described by the same formula which once applied at
Planck's time. The only thing is that the parameters describing
our universe become quantized.

The underling equations describing the universe stem from  Mach's
principle represented by the equation $GM=Rc^2$, where $M$ and $R$
are the mass and radius of the universe. We have found that the
present state of the universe is indeed a  manifestation of its
quantum nature.
\section{\textcolor{blue}{The model}}
The Planck's mass is defined as
\begin{equation}\label{}
m_{\rm P}=\sqrt{\frac{\hbar c}{G}}\ .
\end{equation}
from which one can write
\begin{equation}
\hbar =\frac{Gm^2_{\rm P}}{c}\ .
\end{equation}
We assume here that for any stable bound gravitational (cosmic)
system there exists a cosmic Planck's constant analogue defined
as\footnote{\textcolor{blue}{the subscript `c' hereafter refers to
the cosmic value of the quantity.}},
\begin{equation}\label{}
\hbar_c=\frac{GM^2_{\rm P}}{c}\ ,
\end{equation}
where $M_{\rm P}$ is the Planck's mass of the corresponding cosmic
system. The vacuum (quantum) energy density due to quantum
fluctuation is written, as shown by [8], to be
\begin{equation}\label{}
\rho^{\rm v}=\frac{\Lambda c^2}{8\pi G}=\frac{c^5}{G^2\hbar },
\end{equation}
and consequently one defines a cosmic vacuum density as
\begin{equation}\label{}
\rho_c^{\rm v}=\frac{c^5}{G^2\hbar_c }\ .
\end{equation}
 We obtain from eq.(4) the relation
\begin{equation}\label{}
\hbar= \frac{c^3}{G\Lambda}\ ,
\end{equation}
and for a cosmic system one proposes
\begin{equation}\label{}
\hbar_c=\frac{c^3}{G\Lambda_c}\ .
\end{equation}
Moreover, from a dimensional point of view, one expects the cosmic
Planck's constant to embody quantities having to do with gravity,
(viz.,
 $M_{\rm P}, \ G$ and $\Lambda_c$). Such a
combination is manifested in the form
\begin{equation}\label{}
\hbar_c= \frac{
 G^{\frac{1}{2}} M_{\rm P}^{ \frac{3}{2}  }}{\Lambda_c^{\frac{1}{4}}}\
 ,
\end{equation}
that is valid for a cosmic gravitationally bound system.
\section{\textcolor{blue}{Cosmic and quantum correspondences}}
Eq.(1) can be used to eliminate any mass dependence in the
physical quantities. Hence, one can describe our universe in terms
of the fundamental constants ($G, c, \hbar, \Lambda $) only. For
such a case the physical laws become scale independent and hold
for all systems; microscopic as well as macroscopic. It is worth
to mention that relation
\begin{equation}\label{}
\hbar\ \Lambda=\hbar_c\ \Lambda_c=\frac{c^3}{G}\ , \ {\rm or}
\qquad \rho^v \ \hbar=\rho^v_c\ \hbar_c=\frac{c^5}{G^2}\
\end{equation}
 holds throughout the cosmic expansion. The quantity
$\hbar\Lambda$ represents a rate of formation (creation) of mater.
This rate, which is constant, is equal to $\sim 10^{35}\rm kg/s$.
With this rate the universe must have acquired its present mass
($\sim 10^{53}$ kg) during a time of $10^{17}\rm sec=10^{10}\rm
years$. This indeed coincides with the present condition of the
universe. This is apparent from the fact that $\Lambda$ was so big
in the beginning and decreased to a very small value today at the
expense of creating our present universe. Thus any deviation from
its value in the beginning and its present value, the present
universe (with its current state of being) would not have been
brought. Hence, a non-zero cosmological constant is needed for the
genesis of the universe. We observe that we have used almost all
of vacuum reserve and what is left would not be enough to have
remarkable effect on cosmic expansion. And the cosmic acceleration
which we have now observed can be interpreted as due to the fact
that our universe is very critical and that if mass is generated,
which counteracts the expansion, makes the universe to accelerate
in order to maintain its critical status so that matter (gravity)
should not overwhelm.

For a Machian universe one has
\begin{equation}\label{}
G\rho\sim H^2\ ,
\end{equation}
with $H$ being the Hubble's constant. With the help of eq.(4), the
above equation yields
\begin{equation}\label{}
\hbar\ H^2=\hbar_c\ H_c^2=\frac{c^5}{G}\ .
\end{equation}
This represent the rate of transformation of energy (power).
However, this rate is constant and should be considered as
representing a maximal power, which does not depend on a
particular system.\\
From a purely dimensional argument one can construct a quantum
acceleration  from the set of fundamental constants ($c, \hbar,
G$) of the form\footnote{\textcolor{blue}{see ref. [17] for a
different postulate.}}
\begin{equation}\label{}
a=\left(\frac{c^7}{G\hbar}\right)^{\frac{1}{2}}
\end{equation}
to be valid at Planck's time, and according to our hypothesis, an
analogous acceleration of the form
\begin{equation}\label{}
a_c=\left(\frac{c^7}{G\hbar_c}\right)^{\frac{1}{2}}
\end{equation}
to be valid at cosmic scale. Comparison between eqs.(9) and (11)
shows that one can write the
relation\footnote{\textcolor{blue}{The relation that
$\Lambda\propto H^2$ is found to be interesting (Arbab, A.I.,
2003, Class. Quantum Grav. \textbf{20}, 93)}}
\begin{equation}\label{}
H^2=c^2 \Lambda\ ,
\end{equation}
so that eq.(12) takes the simple form
\begin{equation}\label{}
a=cH\ , \ {\rm or}\qquad a=c^2\sqrt{\Lambda}
\end{equation}
and eq.(13) thus becomes
\begin{equation}\label{}
a_c=cH_c\ , \ {\rm or}\qquad a_c=c^2\sqrt{\Lambda_c}
\end{equation}
 Apparently, one can write this acceleration in different
forms (involving $c, \hbar, G, \Lambda,H$ only) owing to the
relationship between the different fundamental constants.
 We observe that at Planck's time the acceleration was very enormous amounting to $\sim
 10^{51}\rm m\ s^{-2}$, and then evolving to a vanishingly small
 value for the whole universe ($\hbar_c\sim 10^{87}\rm J\ sec$) of $10^{-10}\rm m\
 s^{-2}$. A similar value of $a_c$ today is found by Milgrom\footnote{\textcolor{blue}{Milgrom, M, 1983, ApJ \textbf{270}, 365.}} with
 a different argument. However, this appears in Milgrom theory as a surprise, but we provide here a natural
justification. This
 acceleration is independent of the mass of the object and is a
 universal constant, which every body should manifest. It is a
 characteristic of the present era of cosmic evolution and has to
 do with the existence of the vacuum.
 We expect this acceleration to be prominent (and detectable) at the present time with the use
 of sophisticated tools.
 Thus,  one would expect that our universe today will be filled with such a quantum
 relic(residue).

One asserts that our universe is having critical conditions at the
time. Such conditions have to be satisfied at all stages. Thus our
universe appears to exhibit its dynamic as it evolves from one
state to another. The universe evolves so it satisfies its
critical. This requires some conspiracy among the different
constants making the universe viable. This conspiracy is the
driving power for the universe. We thus remark that the universe
has a maximal power, maximum rate of creation of matter, etc.

The enormous value of the present Planck's constant ($\sim
10^{87}\rm J\ s.$) helps understand why the entropy of the present
universe is so huge. So if we had set $\Lambda$ to zero then the
universe would have become filled with radiation rather than with
matter, as this implies $\hbar_c\rightarrow \infty$. Hence,  this
setting resolve the two (cosmological constant and entropy)
problems simultaneously.

It is believed that the universe satisfies Mach's relation hat
\begin{equation}\label{}
GM=Rc^2\ ,
\end{equation}
where $M$ and $R$ are the mass and radius of the universe,
respectively. This principle is said to have played an important
role in forming Einstein's general relativity. We see that in the
evolving universe Planck's mass, Planck's constant, the
cosmological and the gravitational constants must evolve with time
to satisfy the beauty of the simplicity of our universe. As Albert
Einstein said  ``The most incomprehensible thing about the
universe it is comprehensible" is due to the fact that our
universe is so simple.
\section{\textcolor{blue}{Some numerical estimates}}
Eqs.(3), (5), (7), (8) and (17) are our basic equations for
describing our universe. In order the universe to satisfy eq.(17)
at all epochs, $R$ and $M$ has to evolve with time (from Planck's
time to the present).

In a recent work, we have shown that for this relation to hold
during the radiation epoch, one must have
\begin{equation}\label{}
G\propto t^2\ ,\qquad M\propto t^{-1}\ ,
\end{equation}
and
\begin{equation}\label{}
\Lambda\propto t^{-2}\ .
\end{equation}

We observe that in the early universe the ordinary Planck's
constant ($\hbar$) does not change with time in the early
universe. For this reason the cosmic eqs.(2) and (3) are the same,
i.e., $\hbar=\hbar_c$. By bound system we mean the stars, the
galaxies and the universe. It has been shown that those system
have definite cosmic Planck's constants. It is also shown that the
cosmic Planck's mass of these bound systems (the galaxies and the
present universe) are found to be $\sim 10^{68}\rm J.s$ and $\sim
10^{87}\rm J.s$, respectively from different perspectives [9, 10,
11, 12]. It is also found by Capozziello \emph{et al.} [13] that
Planck's constant for stars is $\sim 10^{52}$ J. s. This would
mean, according to eq.(3), that the Planck's mass for stars system
is $M_{\rm P}\sim 10^{30}$ kg. White dwarf (WD) are
gravitationally bound system having a mass of $1.6\times
10^{30}\rm\ kg$ and density of $1.5\times 10^{9}\rm\ kg m^{-3}$
and extending over a distance of $10^7 \rm m.$ We will see in a
moment, with some scrutiny,  that these characteristics coincide
with our results if one takes for this system  the above values
(viz., $M_{\rm P}\sim 10^{30}\rm\ kg$ and $\hbar_c\sim 10^{52}\rm\
J\ s$). We find from eq.(7) or (8) that $\Lambda_c\sim 10^{-17}\rm
m^{-2}$, which corresponds to a distance scale of $\sim 10^{8}\rm\
m$. Moreover, eq.(5) yields $\rho_v\sim 10^{10}\rm kg m^{-3}$. We
thus see the contribution of the vacuum energy density for such a
system is comparable to the density for the white dwarf (in which
quantum pressure takes over in stopping the overwhelming gravity).

We turn now to calculate the present vacuum energy density of the
universe described by the cosmic numbers, i.e., $\hbar_c$ and
$M_{\rm P}$. The present (denoted by `0') vacuum energy of the
universe, according to eq.(5) will be
\begin{equation}\label{}
\rho^0_{\rm v}\sim 10^{-29}\rm\ g \ cm^{-3}\ ,
\end{equation}
which is of the same order of magnitude of the presently observed
density of the universe. The Compton wavelength of the universe is
given by
\begin{equation}\label{}
\lambda_c=\frac{h_c}{Mc}\sim 10^{26}\ \rm\ m,
\end{equation}
which coincides with the present radius of the universe. This
shows that our universe is indeed a quantum system. For instance,
for a black hole the Schwarzschild radius ($R_s$) given by
$R_s=\frac{2GM}{c^2}$. With the aid of eqs.(3) and (21) this is
transformed into
\begin{equation}\label{}
R_s=\frac{h_c}{Mc},
\end{equation}
which implies that whenever the physical dimension of an object
become of the order of  Compton wavelength quantum effects become
predominant. We thus conclude that the Schwarzschild radius is (or
of the order of) the `cosmic' Compton wavelength. \\ As we have
conjectured before that there is a quantum nature associated with
planets (namely the Earth) of the order of $10^{34}\rm\ J s$ [14],
one finds from eq.(7)
\begin{equation}\label{}
\Lambda_c\sim 10^6 \rm\ \  m^{-2}\ ,
\end{equation}
which corresponds to a millimeter scale, i.e.,
$\frac{1}{\sqrt{\Lambda_c}}\sim 10^{-3}\rm \ m\ .$  This is
nothing but the Schwarzschild radius for the Earth. We remark that
eq.(13) is also consistent with eq.(8) if we wite it to give
$\Lambda_c$.

In terms of the present cosmological constant one gets
\begin{equation}\label{}
\Lambda_0=\frac{c^3}{\hbar_c G_0}\sim 10^{-52}\rm m^{-2}\ ,
\end{equation}
in comparison with the Planckian value
\begin{equation}\label{}
\Lambda_{\rm P}=\frac{c^3}{\hbar G_{\rm P}}\sim 10^{69}\rm m^{-2}\
,
\end{equation}
where we have shown earlier that $G_{\rm P}=G_0$ [12]. In order to
unify gravity with the other three forces, the gravitational
constant has to be a running coupling constant, and during some
time blows up to the scale of the other three forces. It is
expected that gravity get unified at Planck's time. But we have
shown earlier that this is not the case [12]. We expected the real
unification to hold at a rather lower energy scale.

We thus see that $\Lambda$ is a quantized parameter and its value
for a gravitationally bound system can be determined. Therefore,
one gets
\begin{equation}\label{}
\frac{\Lambda_0}{\Lambda_{\rm P}}\sim 10^{-120}\ .
\end{equation}
We, therefore, should be surprised by the vanishingly small value
of the cosmological constant at the present time which follows
from eqs.(24) and (25). What has happened is that the universe is
still describe by its same equations as before and the evolution
of the universe has scaled up some quantifies and scaled down
others. Notice that the speed of light remains constant during all
expansion epochs of the universe.

We see from eq.(17) that the present radius of the universe should
be
\begin{equation}\label{}
R_0=\frac{G_0M_0}{c^2}\sim 10^{26}\rm\ m\ ,
\end{equation}
which matches the observed value. The same equation applies to
Planck's length, namely
\begin{equation}\label{}
R_{\rm P}=\frac{G_{\rm P}M_{\rm P}}{c^2}\sim 10^{-35}\rm\ m\ .
\end{equation}
We thus see,  the apparently complicated universe, how simple it
is. This also shows that how fundamental eq.(17) is, in addition
to how quantum our universe is.

De Sabbata and Sivaram [15] related the the temperature ($T$) to
the curvature ($\kappa$) and showed that
\begin{equation}\label{}
T\propto\sqrt{\kappa}\ \  \propto t^{-1}\ ,
\end{equation}
and a maximal curvature is given by
\begin{equation}\label{}
 \kappa_{\rm max.}=\frac{c^3}{\hbar G}
\end{equation}
 This equation, when compared with
eq.(6), yields
\begin{equation}\label{}
\Lambda\sim \kappa_{\rm max.}\ .
\end{equation}
This result provides another meaning to the cosmological constant
by  relating it to the maximum curvature of the space-time. Hence,
the present smallness of the universe is due to the near-flatness
of our present universe. From eqs.(18), (19) and (30) we see that
the vacuum energy relaxes according to the Planck's law.
Therefore, one must expect a vacuum background filling our
universe at the present epoch having a temperature
\begin{equation}\label{}
T^0_{\rm P}\sim 10^{-29}\rm K\ .
\end{equation}
Thus,  this fluid rolled from a very high value at Planck's time
of with $10^{32}\rm K$ to $10^{-29}\rm K$. This is manifested
differently in the smallness of the present value of the
cosmological constant describing this vacuum. Therefore, the
cosmological constant problem is no longer a puzzle, but defines a
physical that is related to the very nature of our universe. This
small value defines a minimum temperature that any body can
assume. We remark that in the present formalism the hierarchial
nature of the universe is related very much to the quantum
(Planckian) nature. It therefore finds it natural justification;
not as has been justified by the anthropic principle, which often
adopted by cosmologists. Moreover, the quantum states that the
universe can occupy is finite, and those are characterized by
these large cosmological numbers. Equation (32) can be read as
representing the temperature of a quantum fluid relaxing toady and
having such a temperature, or a mass of $10^{-65}\rm g$. If one
relates this mass to  the mass of a long range mediator (possibly
the graviton), then one would say that gravity has a limiting
range and is not infinite, as has long been understood. Or
alternatively, one may interpret this  as that gravity is not
transmitted at the speed of light (c)!
\\
One can arrive at eq.(32)
 by applying the relation
\begin{equation}\label{}
E=m_{\rm P}c^2=kT\ ,
\end{equation}
where $k$ is the Boltzmann constant, as valid at Planck's time.
Using eq.(1) one finds
\begin{equation}\label{}
T=\sqrt{\frac{c^5\hbar}{k^2G}}\ .
\end{equation}
Using eq.(12) one obtains the relation
\begin{equation}\label{}
T=\frac{\hbar a}{ck}\ ,
\end{equation}
a relation connecting the maximal acceleration with the maximal
temperature. Such a relation ($T=\frac{\hbar\ a}{2\pi c k}$) is
proposed by Unruh  [16] that quantum particles should emit thermal
radiation when they are accelerated. According to this proposition
a particle undergoing a constant acceleration would be embedded in
a heat bath at a temperature given by the above equation.
\\
Again, according to our hypothesis the equation
\begin{equation}\label{}
T_c=\frac{\hbar_c}{ck}a_c\ ,
\end{equation}
is valid for cosmic system. We see that eq.(34) yields the
universal values
\begin{equation}\label{}
    a_c\sim 10^{-10}\rm m\ s^{-2},\qquad T_c\sim 10^{-29}\rm K,
\end{equation}
at the present time ($h_c\sim 10^{87}\rm J s$),  and eq.(35)
yields
\begin{equation}\label{}
    a\sim 10^{51}\rm m\ s^{-2},\qquad T\sim 10^{32}\rm K,
\end{equation}
at Planck's time. These findings agree with the earlier values. We
remark here that eq.(33) is the well known formula for the
Hagedorn temperature for elementary particles like pions [18].

 It has been shown by [19] that the
maximal tension in general relativity leads to a relation with the
strings tension (string coupling constant ($\alpha'$, or Regge
slope parameter) given by\footnote{\textcolor{blue}{Since the
maximal force is independent of $\hbar$ one is inclined to
interpret this as that the maximum force in the universe can not
come from any force except gravitational. This maximal force is
the gravitational force exerted by the whole universe. It is a
conserved quantity; its vale at Planck's time is same as to its
value today.}}
\begin{equation}\label{}
\alpha'=\frac{G}{c^4}\ ,
\end{equation}
which is independent of $\hbar$. One, however, can further relate
this to the maximal temperature by the relation
\begin{equation}\label{}
T=\frac{\sqrt{c\hbar}}{k}\frac{1}{\sqrt{\alpha'}} \ .
\end{equation}
However, string theory introduces a temperature scale (known as
Hagedorn temperature) given by [20]
\begin{equation}\label{}
T_{\rm Hagedorn}=\frac{\sqrt{c\hbar}}{4\pi
k}\frac{1}{\sqrt{\alpha'}} \ .
\end{equation}
Consequently one may incline to consider the Hagedorn temperature
as representing a minimal (maximal) temperature. Once again, we
will assume that the cosmic analogue of this equation to exist. We
emphasize here the fact that our universe had  maximal (or
minimal) quantities when it was born and evolving into minimal (or
maximal) quantities by now. This recalls one with some principle
that the universe should respect. This is in essence the duality
principle endowed and reflected by string theories. We have seen
that the maximal temperature occurred at Planck's time and a
minimal one at the present time. Similarly one can view the
present universal Planck's constat ($\hbar_c$) as representing a
maximal value for Planck's constant and $\hbar$ (the ordinary
Planck's constant) as a minimal value for Planck's constant. The
maximal acceleration occurred at Planck's time to be $\sim
10^{51}\rm m\ s^{-2}$ and the minimal acceleration occurring at
the present time. Despite these dualities, which seems at odds,
our physical formulae do respect it. It is worth mentioning that
within this formalism one avoids the occurrence of singularities
(at the beginning or at the end) in our universe.

For completeness we define the Planckian magnetic field density,
electric field, inductance and capacitance  as
\begin{equation}\label{}
 B_{\rm Pl}=\left(\frac{c^5}{4\pi \epsilon_0\hbar
 G^2}\right)^\frac{1}{2},
\end{equation}
\begin{equation}\label{}
 E_{\rm Pl}=\left(\frac{c^7}{4\pi \epsilon_0\hbar
 G^2}\right)^\frac{1}{2},
\end{equation}
\begin{equation}\label{}
 L_{\rm Pl}=\frac{1}{4\pi \epsilon_0}\left(\frac{\hbar
 G}{c^7}\right)^\frac{1}{2},
\end{equation}
\begin{equation}\label{}
 C_{\rm Pl}=4\pi \epsilon_0\left(\frac{\hbar
 G^2}{c^3}\right)^\frac{1}{2}\ ,
\end{equation}
where $\epsilon_0$ is the permittivity of free space. According to
our hypothesis the present (universal) values of the above
quantities are
\begin{equation}\label{}
B_{\rm Pl}\sim 10^{-8}\rm T,\ E_{\rm Pl}\sim 1\rm V/m,\ L_{\rm
Pl}\sim 10^{19}\rm H, \ C_{\rm Pl}\sim 10^{16}\rm F\ ,
\end{equation}
representing the relic (residue) values of the Planckian time. We
remark that one can related the above quantities to Planck density
and acceleration. We found, today, that the  Planckian capacitance
and inductance reach their maximal values, while the electric
field and magnetic field density relax to their minimal values. If
one calculate these values for galactic ($\hbar_c\sim 10^{68}\rm J
s)$ and solar ($\hbar_c\sim 10^{52}\rm J s)$ levels, one
respectively arrives at
\begin{equation}\label{}
B_{\rm Pl}\sim 10^{2}\rm T,\ E_{\rm Pl}\sim 10^6\rm V/m,\ L_{\rm
Pl}\sim 10^{9}\rm H, \ C_{\rm Pl}\sim 10^{6}\rm F\ ;
\end{equation}
and
\begin{equation}\label{}
B_{\rm Pl}\sim 10^{10}\rm T,\ E_{\rm Pl}\sim 10^{18}\rm V/m,\
L_{\rm Pl}\sim 10\rm H, \ C_{\rm Pl}\sim 10^{-2}\rm F\ .
\end{equation}
If one believes in the gravitional-electrical symmetry we come up
with the definition of a cosmic-magnetic moment defined as
\begin{equation}
\mu^c_B=\frac{q\hbar_c}{2M}\sim\sqrt{\left(\frac{G}{k}\right)}\ \
\hbar_c\ ,
\end{equation}
where we have used the gravitational charge ($q$) of a cosmic body
as $kq^2=GM^2$ (where $k=\frac{1}{4\pi \epsilon_0}$). With this
prescription one reveals that the present universe has a cosmic
magnetic moment $\sim 10^{68}\rm J/T$ so that it is now dominated
by gravito-magnetic energy  $ E=\mu^c_B B_{\rm Pl}\sim
10^{78}\times 10^{-8}=10^{70}\rm J$.  Since a galaxy has a
gravito-magnetic moment of $\sim 10^{58}\rm J/T$,  one finds this
gravito-magnetic energy to be $\sim 10^{60}\rm J$. This would
imply that the universe contains some $10^{11}$ galaxies. For the
universe at Planck's time its magnetic moment would be $\sim
10^{-44}\rm J/T$ so that its gravito-magnetic energy should be
$\sim 10^9\rm J$.

With the same token, one defines the gravito-electric dipole
moment as $\mu^c_E=qd=\sqrt{\left(\frac{G}{k}\right)}Md$, where
$d$ is the distance between the dipoles. By virtue of this
equation, the gravito-electric energy of the universe today would
be $E=\mu^c_EE_{\rm Pl}= 10^{-10}\times 10^{53}\times
10^{26}\times1 \sim10^{70}\rm J$. The gravito-electric energy of
the universe at Planck's time would be $E=\mu^c_EE_{\rm Pl}=
10^{-10}\times 10^{-8}\times 10^{-35}\times 10^{61}\sim 10^9\rm
J$. These calculations give very consistent and reasonable values
that are presently known to describe our universe. Apparently, the
energy of the universe today is dominated equally by
electromagnetic as well as gravitational interactions. Despite
this fact no one today does think that such an electromagnetic
contribution exist. This unseen contribution might solve the
missing energy of the universe. Thus, the existence of dark energy
(quintessence) that resolve this missing mass might not be
necessary; it might be the second blunder that contemporary
cosmologists have made after Einstein to justify the present
cosmic acceleration! We thus see that the electromagnetic
interactions are present even at cosmic level, but not apparent.
They are embedded in gravity. We remark that an acceleration of an
order of $10^{-10}\ ms^{-2}$ can be interpreted as due to a unit
mass of gravitational
 charge $q=\sqrt{4\pi \epsilon_0G}$ moving in a
residual electric field, as given by eq.(46). We notice this is
the same as that obtained in eqs.(13) and (16). We would expect to
come up with these quantities in our search at these cosmic levels
outlined above. These relics (residues) should be found at the
present time. Thus any experimental endeavor to find these
\emph{backgrounds} physical quantities will be very interesting.

The maximal and minimal cosmological quantities are summarize in
table 1.
\section*{\textcolor{blue}{Conclusion}}
We have seen that our universe  is an indeed evolving quantum
system. It is governed by the Mach relation and the vacuum
(quantum) energy density that has been continually contributing to
the total energy density of the universe. We have seen that the
different mass scales we observe today is a manifestation of the
quantum nature of our universe. Thus our universe is really
quantum evolving system. This simple description outlined in this
letter solves the cosmological constant problems and the
hierarchial problems upsetting our standard model of cosmology.
The idea that our universe is a quantum system fits remarkably
well with the present condition of the universe. We have shown
that one can describe the status of the universe by a set of
fundamental constants  (eg., $c, \hbar, G, \Lambda, H$) only. This
would require a cooperation (conspiracy) between these constants
in order to satisfy this eternal picture of our universe. We
believe that present description of the universe we postulated,
allows for a conformal representing of the basic laws underlying
these formalism. This is because we have not considered any mass
scale in our basic equations. This makes our universe looks quite
simple and understandable. We remark that the present formalism,
albeit non-formal, it shed a lot of lights on the real formulation
of quantum gravity that every one awaits its advent. We have
provided a semi-quantum approximation for a classical system.
universe. By writing our basic formulae of the universe in terms
of the fundamental constants ($c, \hbar, G, \Lambda, H$), we have
in principle considered a unification of all interactions
(relativity, quantum, gravity, vacuum and cosmology), but in a
rather non formal approach. Though our treatment of quantization
of gravitational system is crude it however gives a flavor of the
subject. Within this framework all phenomena appearing in the
universe can be seen as a realization of this unification. We
believe that a real quantization of gravity should include the
electromagnetic interactions that are present even at cosmic
level.
\section*{\textcolor{blue}{Acknowledgements}}
I would like to thank the abdus salam International Cenetre for
Theoretical Physics (ICTP) for hospitality and  Comboni College
for providing a research support for this work.
\\
\section*{\textcolor{blue}{References}}
1. S. M. Carroll, \emph{Living Rev.Rel.} \textbf{4}, 1 (2001) \\
2. S. M. Carroll, W.H. Press, and E.L. Turner, \emph{Ann. Rev. Astron. Astrophys.} \textbf{30}, 499 (1992).\\
3. S. Weinberg,  \emph{Rev. Mod. Phys.} \textbf{61}, 1 (1989).\\
4. T. Padmanabhan,  \emph{Phys.Rept}. \textbf{380}, 235 (2003).\\
5. V. Sahni and A. A. Starobinsky, \emph{Int.J. Mod. Phys.}\textbf{D9}, 373 (2000).\\
6. J. M. Overduin, F.I. Cooperstock , \emph{Phys.Rev}.\textbf{D58} 043506 (1998).\\
7. J. A. Belinchon \emph{Int.J.Theor.Phys.} \textbf{39} 1669 (2000).\\
8. J. A. S Lima and J. C. Carvalho, {\it Gen. Rel. Gravit.} \textbf{26}, 909 (1994).\\
9. M. Dersarkissian, \emph{Nouvo Cimento lett.} \textbf{40}, 390
      (1984)\textbf{;} \emph{Nouvo Cimento lett.} \textbf{43}, 274 (1985).\\
10. C. Massa, \emph{Lett Nouvo Cimento}. \textbf{44}, 671 (1985).\\
11. P. Caldirola, N. Pavsic, and E. Recami, \emph{Nouvo Cimento.} \textbf{B48}, 205 (1978).\\
12. A. I. Arbab,  {\it Spacetime \& Substance Journal.} \textbf{7}, 51 (2001)\textbf{;} astro-ph/9911311.\\
13.  S. Capozziello, S.  Martino, S. Siena, and F. Illuminati, gr-qc/9901042.\\
14. A. I. Arbab, \emph{Spacetime \& Substance} \textbf{2}, (7), 55 (2001). \\
15. V. de Sabbata and C. Sivaram, \emph{Astrophys. Space.
Science.}\textbf{158}, 947 (1989)\\
16. W. G. Unruh, \emph{Phys. Rev.}, D\textbf{14}, 870 (1976).\\
17. E. R, Caianiello,  \emph{Lett. Nouvo Cimento}.\textbf{41}, 370
(1984); \textbf{32}, 65 (1981).\\
18. B. G. Sidharth, physics/0302054\\
19. G. W, Gibbons, hep-th/0210109.\\
20. D.L. Wiltshire, gr-qc/0101003\\
\newpage
\begin{table}[htb!]
\caption{The values (in order of magnitudes) of the maximal and
minimal physical quantities in the universe} \vspace{0.4cm}
\begin{center}
\begin{tabular}{|r|r|r|r|}
\hline \textcolor[rgb]{1.00,0.00,0.00}{Planckian quantity}  & \textcolor[rgb]{1.00,0.00,0.00}{Unit} & \textcolor[rgb]{1.00,0.00,0.00}{Maximal value} & \textcolor[rgb]{0.98,0.00,0.00}{Minimal value}\\
\hline
\textcolor[rgb]{0.00,0.00,1.00}{Planck's constant} & \rm J\ s &  $10^{87}$ & $10^{-34}$ \\
\hline
\textcolor[rgb]{0.00,0.00,1.00}{Acceleration }& $\rm m\ s^{-2}$ & $10^{51}$ & $10^{-10}$\\
\hline \textcolor[rgb]{0.00,0.00,1.00}{Mass }& $\rm kg$ & $10^{53}$ & $10^{-8}$  \\
\hline \textcolor[rgb]{0.00,0.00,1.00}{Time} & $\rm sec$ & $10^{17}$ & $10^{-44}$ \\
\hline \textcolor[rgb]{0.00,0.00,1.00}{Density} & $\rm kg\ m^{-3}$ & $10^{96}$ & $10^{-26}$\\
\hline \textcolor[rgb]{0.00,0.00,1.00}{Temperature} & $\rm K$ & $10^{32}$& $10^{-29}$\\
\hline \textcolor[rgb]{0.00,0.00,1.00}{Cosmological constant} & $\rm m^{-2}$ & $10^{69}$ & $10^{-52}$\\
\hline \textcolor[rgb]{0.00,0.00,1.00}{Inductance} & $\rm H$ & $10^{19}$ & $10^{-42}$\\
\hline \textcolor[rgb]{0.00,0.00,1.00}{Capacitance} & $\rm F$ & $10^{16}$ & $10^{-45}$\\
\hline \textcolor[rgb]{0.00,0.00,1.00}{Electric field} & $\rm V/m$ & $10^{61}$ & $10^{0}$\\
\hline \textcolor[rgb]{0.00,0.00,1.00}{Magnetic field density} & $\rm T$ & $10^{53}$ & $10^{-8}$\\
\hline
\end{tabular}
\end{center}
\end{table}
We observe that these quantities are either 61 or 122 orders of
magnitude when compared between Planck era and today! Without
these finetunings  our universe would not have remained for
billion years.
\end{document}